\documentclass[aps, prd, twocolumn, lengthcheck, superscriptaddress, showpacs, nofootinbib]{revtex4-1}

\usepackage{epsfig}
\usepackage[usenames]{color}
\usepackage{graphicx}
\usepackage{amsmath}
\usepackage{epstopdf}

\newcommand\sect[1]{\emph{#1}---}
\def\bi{\bibitem}

\def\la{\langle}\def\ra{\rangle}
\def\be{\begin{eqnarray}}\def\ee{\end{eqnarray}}
\def\lsim{\mathrel{\rlap{\lower3pt\hbox{\hskip1pt$\sim$}}
     \raise1pt\hbox{$<$}}} 
\def\gsim{\mathrel{\rlap{\lower3pt\hbox{\hskip1pt$\sim$}}
     \raise1pt\hbox{$>$}}} 
\def\del{\partial}
\def\V{\cal V}\def\O{\cal O}

\allowdisplaybreaks

\begin{document}

\title{Resolving  the  Quenched  ${g_A}$ Puzzle \\  in Nuclei and Nuclear Matter}

\author{Mannque Rho}
\email{mannque.rho@ipht.fr}
\affiliation{Institut de Physique Th\'eorique,
	CEA Saclay, 91191 Gif-sur-Yvette c\'edex, France }
\date{\today}

\begin{abstract}
The long-standing puzzle of the quenched $g_A$ in nuclei -- and in dense matter -- is shown to have  a simple resolution in a scale-symmetric HLS chiral Lagrangian at the Fermi-liquid fixed point. This resolution exposes scale-chiral symmetry, hidden in QCD in the vacuum, emerging in nuclear matter from low density to high compact-star density.  It has important implications on ``first principles" approaches to nuclear physics, such as the role of multi-body exchange currents in Gamow-Teller matrix elements in nuclei and in neutrinoless double $\beta$ decays for going Beyond the Standard Model.
 
\end{abstract}

\maketitle

\sect{\bf Introduction}
%
%
There is a long-standing ``mystery" lasting more than four decades as to why the Gamow-Teller transition in shell model in nuclei requires a universal quenching factor $\sim 0.75$ multiplying the axial coupling constant $g_A$ measured in neutron decay, which would make the effective axial-vector  coupling constant $g_A^{\rm eff}\approx 1$. What was striking then -- and is more so now -- is that the resulting $g_A$ is surprisingly close to 1 in light and medium nuclei as updated in  the recent review~\cite{review-gA}.  This prompted Denys Wilkinson from early 1970s~\cite{wilkinson}  -- and others afterwards up to today since then -- to inquire whether this is not associated with something intrinsically tied to a basic property of QCD in nuclear medium. While the conserved vector current implies that the vector coupling constant $g_V=1$ in and out of medium, the conserved axial current --- in the chiral limit -- does not imply that $g_A=1$. This is now understood as that the axial symmetry is a ``hidden" symmetry unlike the  vector symmetry which is unhidden.

There is no known answer as to whether the apparent nuclear $g_A^{\rm eff}\approx 1$ is just a coincidence and if not, what it means. In this paper, we provide a possible  explanation built on an old resolution of the problem.

To see what the problem is, consider the celebrated Adler-Weisberger sum rule that follows from the current algebras of chiral symmetry~\cite{adler,weisberger},
$g_A^2=1+ f_\pi^2\frac{2}{\pi}\int^\infty_{m_N+m_\pi} \frac{WdW}{W^2-m_N^2} \big[\sigma^{\pi^+ p} (W)-\sigma^{\pi^- p} (W)\big]$.  Applying the sum rule for the proton to a nucleus $A$ treated as an ``elementary particle"~\cite{kim-primakoff}, one notes that  $g_A\to 1$ if either $f_\pi\to 0$ or the integral over the difference of $\pi^\pm A$ scattering vanishes. There is nothing that suggests that the second term should vanish unless $f_\pi \to 0$. It is believed that the pion decay constant will indeed go to zero -- in the chiral limit -- at some high density independently of how the second term behaves.  In finite nuclei, however,  in the vicinity of  the equilibrium nuclear matter density $n_0$, the in-medium $f_\pi^\ast $  could drop, at most $\sim 20\%$, from the free space value  and there  is  no known reason why the integral is to vanish. Thus the puzzle of the effective $g_A^\ast$ tending to go near 1 in a wide range of nuclei.

This issue got highlighted recently by a remarkable ``work-of-the-art" computation of Gamow-Teller transitions in light and medium-nuclei
~\cite{firstprinciple}. This work combines no-core shell model technique, thereby incorporating what could be, as claimed by the authors of \cite{firstprinciple},``virtually exact" correlations in the nuclear wavefunctions and EFT treatment of strong and weak interactions of the Standard Model.
What  makes this work particularly significant  is that it  is focused on the super-allowed Gamow-Teller decay of the doubly magic $^{100}$Sn nucleus which exhibits the strongest Gamow-Teller strength so far measured in nuclei~\cite{Sn100},  an ideal system for large-scale calculation that can take into account a large number of particle-hole correlations.   It predicts 
the quenching factor $q=0.73-0.85$ for $^{100}$Sn, agreeing with  $q_{B_{\rm GT,ESPM}}=0.75(2)$ obtained in  \cite{Sn100}, thereby giving $g_A^{\rm eff}= 0.95 - 1.08$. This calculation, as the title indicates, is heralded as a ``first-principles" resolution of the long-standing puzzle. The crucial ingredient in this ``resolution" is  the two- and three-body weak currents treated 
in what is claimed to be a systematic chiral power expansion in the nuclear EFT that  involves nucleons and pions only, denoted hereon as S$\chi$EFT$_\pi$.

In this paper we show that nucleon particle-nucleon hole correlations involving excitation energies up to but less than the $\Delta N$ mass difference $\sim 300$ MeV should fully -- if not entirely -- account for the {\it universal} quenching factor. Neither multi-body meson-exchange currents nor baryon resonance degrees of freedom, notably, the $\Delta$s, are significantly  involved. The  Lagrangian referred to hereon  as ``$bs$HLS," constructed so as to apply to baryonic matter ranging in density from normal nuclear matter ($n_0$)  to compact-star matter ($\sim (5-7)n_0$)~\cite{MR-review}, connects what is at work for the $g_A$ at low density to the emergence of scale symmetry at high density in compact stars.\footnote{This Lagrangian involving only hadronic degrees of freedom encodes, in terms of a topology change,  what corresponds to the hadron-quark continuity in QCD at a density $\gsim 3n_0$. The cross-over is absolutely essential for the EoS of massive neutron stars, but it affects only indirectly, through nuclear tensor forces, the $g_A$ problem.}  What is highly significant is that  this resolution has far-reaching implications on ``first-principles" approaches to nuclear physics and  the powerful roles of hidden symmetries of QCD in nuclear dynamics. It also has implications on the double Gamow-Teller $\beta$ decay  elements for going Beyond the Standard Model.

\sect{\bf Standard Nuclear Chiral EFT}
To see what is involved in the puzzle and its resolution proposed in \cite{firstprinciple}, let us see what can be amiss with the problem in S$\chi$EFT$_\pi$.
The calculation of \cite{firstprinciple} is based  on the standard chiral EFT, S$\chi$EFT$_\pi$, that is to capture low-energy nonperturbative QCD  by including, as relevant degrees of freedom,  only the pions in addition to the nucleons -- proton and neutron.  Apart from the symmetries etc. required, what is needed for nuclear EFT is the effective cutoff scale $\Lambda_{\rm eff}$ involved in nuclear dynamics. In the usual S$\chi$EFT$_\pi$ calculations practiced in the field as in \cite{firstprinciple}, the cutoff is typically taken $\Lambda_{\rm eff}\sim 400-500$ MeV. This means that  massive degrees of freedom such as the lowest-lying vector mesons, $V = (\rho, \omega)$, are integrated out  as their free-space mass is greater than the cutoff $\Lambda_{\rm eff}$. Furthermore no scalar figures as an explicit degree of freedom.  First the  scalar ``$\sigma$" playing an indispensable role  in the relativistic mean field theories (referred in the literature to as RMFT)
 is of higher mass than $\Lambda_{\rm eff}$, so it is also integrated out. Secondly because it can appear as a resonance at high loop-orders in S$\chi$EFT$_\pi$, its explicit presence risks the danger of double counting.

In contrast, however, the nucleons with the mass $\sim 1$ GeV do figure {\it explicitly} in nuclear EFT for the obvious reason that nucleons are essential. They could be integrated out but then the nucleons would have to be brought in as skyrmions in the Lagrangian. This matter is discussed in \cite{MR-review}. However this round-about procedure is unnecessary because what is involved in low energy nuclear physics are ``soft" fluctuations comparable to soft pions in low-energy strong interactions.  Now what about the $\Delta (3,3)$ resonance which is some $\sim 300$ MeV heavier?  The answer is that it does not enter in the quantum many-body approach adopted here as will become clear.
%

In \cite{firstprinciple}, the nuclear forces ${\cal V}$ and the manny-body currents ${\cal O}$ are considered, respectively,  up to N$^4$LO  and  N$^3$LO  in the given chiral power counting.\footnote{We use the  conventional notation doing the power counting {\it relative} to the leading order in the ``small" expansion (momentum or quark mass) parameter $Q$. Specifically if the leading order (LO) term is of $O(Q^k)$, the subleading term of $O(Q^{k+m})$  for $m> 0$ is denoted as N$^m$LO etc.} With a highly reliable  $\V$
the wavefunctions are {\it precisely calculated} given the powerful quantum many-body technique.

 Now the issue of the many-body currents $\O$ needs  to be addressed.

 It turns out that unlike the vector currents that are quite straightforward the nuclear axial-vector currents turn out to be extremely  subtle.   This has to do with that the axial symmetry is ``hidden."

 In fact it has been known since late 1970s that the space and time components of nuclear axial currents behave quite differently in nuclei and dense baryonic matter. This was evidenced in the current algebras before the advent of QCD in the way ``soft pions" come into two-body exchange currents~\cite{KDR}. In terms of the modern $\chi$EFT parlance, this is almost trivial. However the soft-pion theorems, just as all other soft theorems, be that photon or graviton, have a deep physical implication, ubiquitous in all areas of physics~\cite{soft}.
 %
 %

 A simple heuristic picture of what is going on is as follows.

 Consider the two-body axial currents with one-pion exchange which in the S$\chi$EFT$_\pi$  are dominant in the chiral power counting as we will see precisely below.  Since the axial field ${\cal A}_\mu$ does not couple to the pion exchanged between nucleons, what is involved is the vertex ${\cal A}_\mu +N \to \pi +N$. We can take  the axial field ${\cal A}_\mu$ as a pion. Thus  we are dealing with the process $\pi_{\rm in}+ N\to \pi_{\rm out} +N$ where $\pi_{\rm in}$ stands for the incoming axial field and $\pi_{\rm out}$ is the pion exchanged between two nucleons.

 First consider the case where  $\pi_{\rm in}$  is ``hard" and $\pi_{\rm out}$  is ``soft."  Then according to the double soft theorems~\cite{soft},  the amplitude should be highly {\it suppressed} by Adler's theorem. On the other hand, if both $\pi_{\rm in}$  and $\pi_{\rm out}$  are soft, then the double-soft limit gives an  unsuppressed $\sim O(1)$ amplitude. This is very well known from the old soft-pion theorems, but nowadays this old stuff has become highlighted  because of its fundamental nature in physics~\cite{soft}.

 Kinematics of the virtual  pion in nuclei is not  sharply given, so our argument is at best approximate. But with the axial current identified with a pion, this soft-theorem can be applied to the problem. The pion exchanged between two nucleon favors the process when it is soft, with harder pions suffering from kinematic suppression due to the derivative coupling. Now taking the axial charge operator ${\cal A}_0$ as a soft pion, this then predicts an $O(1)$ one-pion exchange contribution whereas the Gamow-Teller operator ${\cal A}_{\pm}$, being ``hard," leads to a suppressed two-body operator.  This is essentially the ``chiral filter" argument of \cite{KDR}. 

The above chiral filter argument can be given a more rigorous support by using the systematic power counting in S$\chi$EFT$_\pi$ to which we turn.

The power counting for the responses to the electro-weak current,  first worked out in early 1990s and listed completely in a publication in 2003~\cite{parketal}, has  been extensively refined and extended since then as aptly summarized -- with relevant references -- in  \cite{krebs}. What we need for our arguments is essentially all contained in \cite{parketal} that we will follow. We will, however, exploit some of the critical analyses made in \cite{krebs} on consistency in the regularization schemes in going to N$^4$LO.

What follows is a story of two sides of the same coin.

We first look at the time component of the axial current. Here soft pions predominantly enter in the two-body current. The ratio of the two-body soft-pion exchange operator over the one-body operator -- which is $O(Q)$ in the $Q$-power counting -- is $R=$2B/1B$=O(Q^0)$. Thus the leading ``correction" is of the same magnitude as the LO one-body term. The next correction is suppressed by  two chiral orders,  $O(Q^2)$. At this order there are  relativistic and other small corrections  to the single-particle operator as well as two-body terms involving 2$\pi$ exchange etc.  They are expected to be ignorable. Thus the leading two-body term is protected by  the ``chiral filter,"  hence robust.

This prediction has been neatly confirmed in the first-forbidden $A$-to-$B$ nuclear $\beta$ decay  $A(0^\pm)\to B(0^\mp) +e +\nu, \ \ \Delta T=1$
 where the superscripts are the parities. Expressed as $\epsilon={g_A}^{\rm eff}_t/g_A$  in terms of the effective axial coupling constant for the time component to represent the ratio of the total matrix element over the single-particle matrix element,   the prediction for nuclear matter~\cite{KR91} $\epsilon_{theory}=2.0\pm 0.2$ and the experimental measurements made for the transitions in Pb region $A=205 -212$~\cite{warburton} $\epsilon_{exp}=2.01\pm 0.05$  agree  stunningly well.  The theoretical value is estimated at nuclear matter density, but the result is extremely insensitive to density, so Pb can be compared with nuclear matter: The 10\% error bar assigned to  the theory corresponds to the range of density involved from light to heavy nuclei to nuclear matter.  This result is well supported in other processes involving lighter nuclei.
 This perhaps is the most convincing  -- and clear-cut -- evidence for the role of soft pions -- via exchange currents --  in nuclear physics.

 This is the story of one side of the coin.

Now we look at  the Gamow-Teller operator, the space component of the axial current. The situation here  is drastically different. This is because soft pions play practically no role here.  While the one-body Gamow-Teller operator is $O(Q^0)$,   super-allowed -- barring  accidental suppression,    the leading two-body correction with one-pion exchange comes ( \`a la  soft theorems)   strongly suppressed by two chiral orders,  $O(Q^2)$. The ratio is $R=$2B/1B$=O(Q^2)$. This is because the pion entering in the two-body term is ``hard," with its coupling with nucleons requiring, among others, relativistic corrections.  At this order, three-body operators must also enter. Furthermore since the nucleons are inevitably non-relativistic, there can be a plethora of other corrections, notably  the  ``recoil corrections," entering at the same  order. It does not appear from what's discussed in the paper that all these corrections are fully and consistently taken into account in \cite{firstprinciple}. There is no justification to take  some but ignore others as there can be significant cancellations among them. They should also be all essential for axial Ward identities.   To make it even worse, there are also ambiguities in doing regularizations in both $\V$ and $\O$~\cite{krebs}, relevant to the validity in correlating the presence of 2BC with the regularization (cutoff dependence,   a.k.a. ``resolution scale" etc.) that figures importantly in the arguments given in \cite{firstprinciple}.  This large number ($> 11$) of  higher-order terms that cannot be controlled at N$^k$LO for $k\geq 3$ is what is meant by ``chiral-filter unprotected" terms.

This is the story of the other side of the same coin.

In sum,  we conclude that the conclusion of \cite{firstprinciple} -- that the $g_A$ problem is resolved by the  2BC combined with a sophisticated no-core shell model -- is highly questionable if not unfounded. There is no justification to stop at N$^3$LO unless N$^4$LO can be shown to be ignorable, which is at present far from feasible. There can very well be cancelations between different orders as in the case of the Monte Carlo calculations in light nuclei~\cite{wiringa}.

\sect{\bf Scale-Chiral  Hidden Local Symmetric EFT}
We now present an alternative calculation which provides a strong support to the conclusion given above. This calculation relies on the $bs$HLS Lagrangian mentioned above. As reviewed in \cite{MR-review}, it works well without any inconsistency with the presently known properties of nuclear matter as well as of compact-star matter. It is rather involved including a possible hadron-quark continuity at high density. But  it drastically simplifies for densities $\leq n_0$, so is  applicable to the problem concerned. It involves only one parameter, namely, the pion decay constant in medium $f_\pi^\ast$.
%

The Lagrangian $bs$HLS  is constructed with a cutoff put above the vector-meson mass $\sim 700$ MeV, with the vector mesons $\rho$ and $\omega$ brought in as hidden gauge fields and a scalar corresponding to $f_0(500)$, denoted $\chi$ (to be distinguished from $\sigma$ of linear sigma model and also from the scalar in RMFT)  as a (pseudo-)Nambu-Goldstone scalar boson of scale symmetry. Those, in addition to the nucleons, are the relevant degrees of freedom for the given cutoff. The $\Delta$s do not figure. It has the same chiral symmetry as S$\chi$EFT$_\pi$ but goes way beyond it with the hidden symmetries emerging in nuclear medium. There is no need for ``resolution-scale" adjustment.

The power of  the  $bs$HLS Lagrangian applied to nuclear physics is that its bare parameters are endowed with non-perturbative inputs in terms of condensates inherited from QCD at the matching scale between the EFT and QCD. Since the condensates track the vacua, the Lagrangian is endowed with what is called ``intrinsic density dependence (IDD)." The Lagrangian is constructed to apply to a wide range of densities from that of nuclear matter $n_0$ to compact-star matter $\sim (5-7) n_0$.  It possesses hidden  gauge symmetry encoding  chiral symmetry in terms of the vector mesons gauge equivalent  to non-linear sigma model. It encodes the vector manifestation fixed point at which the gauge coupling goes to zero. It contains also a scalar degree of freedom, the dilaton  $\chi$ encoding the scale symmetry of QCD. This symmetry is hidden in the QCD vacuum and shows up in nuclear medium triggered by nuclear matter. The scale symmetry and HLS, treated on the same footing, give rise to scale-chiral symmetry which is taken as the basis of nuclear strong dynamics in lieu of the chiral symmetry alone. The presence of the dilaton as an active degree of freedom makes the theory a lot more powerful -- and potentially much simpler --  than chiral symmetric theory.

The scale symmetry with which the scalar dilaton is associated has a long history and there is still controversy as to whether such a scalar exists in QCD for the number of flavors less than or equal to 3 relevant to nuclear physics. If it exists it must be basically different from such scalar present in conformal window being studied for dialtonic Higgs for large number of $N_f$. We eschew the controversy, referring to \cite{MR-review} for some relevant discussions. Here we are following the scheme developed in \cite{CT} which  we find appropriate for nuclear dynamics. It captures economically the effect of  high loop-orders in S$\chi$EFT$_\pi$ at tree order as is exemplified in particle physics in the process in $K\to 2\pi$ decay~\cite{CT}.





Written schematically, the $bs$HLS Lagrangian is of the form
\be
{\cal L}_{bsHLS}={\cal L}_{inv} (\psi, U, \chi, V_\mu) + {\cal V} (U,\chi, {\cal M})\label{LOSS}
\ee
where the first term is scale-invariant and the second is the dilaton potential that encodes scale-chiral symmetry breaking.  This results from the leading-order scale symmetry (LOSS) approximation found to be appropriate for nuclear dynamics~\cite{MR-review}. In particular it encodes soft theorems. Here $\psi$ is the nucleon field,  $U=e^{2i\pi/f_\pi}$ is the chiral field, $\chi=f_\chi e^{\sigma/f_\chi}$ is the ``conformal compensator field" for the dilaton,  $V_\mu$ is the hidden gauge field. The dilaton potential encodes the vacuum structure putting the system in the Nambu-Goldstone mode of scale-chiral symmetry.  The HLS is assured with hidden gauge covariance put in the Maurer-Cartan 1-forms and can be written down to any power orders.

Given the $bs$HLS Lagrangian, one can formulate a quantum many-body approach to nuclear dynamics using Wilsonian-type renormalization group flow such as, for instance, the $V_{lowk}$-RG.
How this works in compact stars is described in \cite{MR-review}.  It leads to a Landau Fermi-liquid theory formulation of the EoS for dense matter~\cite{fermi-liquid}. What we do is effectively applying the same operation to the $g_A$ problem but in a much simplified form. Renormalization-group decimated to the top of the Fermi sea,  the  quasiparticle making the Gamow-Teller transition can be identified with the pure shell-model transition for $q_{B_{\rm GT,ESPM}}$. We calculate this quantity by applying the Landau Fermi-liquid fixed-point approximation which corresponds to taking the limit $\bar{N}\equiv \frac{k_F}{\Lambda-k_F}\to \infty$~\cite{shankar}.

One can see from the explicit expression of the Lagrangian (\ref{LOSS}) that to the lowest order in scale-chiral expansion,   the Lagrangian has the form of Walecka's linear mean-field model~\cite{walecka}. The major difference from Walecka's model  however is that  the bare parameters,  endowed with the IDDs, are constrained by hidden local symmetry (hence chiral symmetry) and scale symmetry (hence low-energy theorems with the dilaton)  and of course the pion fields included \`a la nonlinear chiral symmetry.
%
Now the most crucial point in our approach is that {\it this $bs$HLS Lagrangian with the IDDs suitably implemented, when treated in the mean-field, is equivalent to the Landau Fermi-liquid fixed point theory.}  Such an  ``equivalence" was suggested a long time ago by Matsui for Walecka's linear RMFT~\cite{matsui} and  was shown even quantitatively  to hold for a Lagrangian of the type of (\ref{LOSS}), including the  thermodynamic consistency~\cite{song}.

To see how the equivalence works,  we consider the nuclear response to the EM field. Here the situation is  straightforward. An illustrative case is the EM orbital current in the mean-field treatment of (\ref{LOSS}) which reproduces precisely Migdal's finite Fermi-liquid formula~\cite{migdal} $
\vec{J}=\frac{\vec{k}}{m_N} \big(\frac{1+\tau_3}{2} +\delta g_l\big)$ with
$\delta g_l = \frac 16 (\tilde{F}_1^\prime -\tilde{F}_1)\tau_3$
where $\tilde{F}_1$ and $\tilde{F}_1^\prime$ are Landau-Migdal interaction parameters expressed in terms of the parameters of $bs$HLS. There are two remarkable results in this formula. First  the orbital current  is given in terms of the vacuum nucleon mass -- instead of the Landau mass $m_L$ -- satisfying the Kohn theorem~\cite{kohn} and the other is that the prediction for the nuclear anomalous  gyromagnetic ratio~\cite{Friman-Rho} -- with the soft-pion theorems playing the crucial role -- $\delta g_l^p (n_0)\simeq 0.21$ agrees with what's measured in the Pb region, $\delta g_l^{\rm proton} = 0.23\pm 0.03$~\cite{schumacher}.

We finally turn  to the $g_A$ problem, which is linked to low-energy theorems in the axial channel that are quite intricate. We focus on the quenching factor $q$ associated with the $B_{\rm GT,ESPM}$ in $^{100}$Sn~\cite{Sn100}.

The quenching factor in the Fermi-liquid fixed point theory is  $q_{L}=g_{A}^{L}/g_A$ where $g_A^{L}$ is what corresponds to the Fermi-liquid fixed point constant that multiples the zero-momentum-transfer matrix element ${\cal M}=(\sum_i \tau_i\sigma_i)_{QP}$ for the quasi-particle on top of the Fermi surface making the GT transition, $M_{GT}=q_L g_A {\cal M}$. This $q^L$ should be compared with the experimental value
$q$ in $^{100}$Sn.

The relevant part of the Lagrangian (\ref{LOSS}) for this problem is
\be
 {\cal L} &=&i\bar{\psi} \gamma^\mu \del_\mu \psi -\frac{\chi}{{f_\chi}}m_N \bar{\psi}\psi +g_A \bar{\psi}\gamma^\mu\gamma_5 \tau_a \psi{\cal A}_{\mu}^a+\cdots
 \label{LAG}
 \ee
 where ${\cal A}_\mu$ is the external axial field.   The key point  to note here  is that the axial response term in the Lagrangian is scale-invariant without dependence on the conformal compensator field $\chi$, whereas the nucleon mass term is linear in $\chi$. This means that embedded in nuclear medium, $g_A$ as a bare parameter is free of  IDDs, whereas the nucleon mass does scale ``intrinsically" as
  $m_N^\ast =\Phi m_N$ with  $\Phi \equiv  f_\chi^\ast/f_\chi$. Due to the locking of the condensates, we can identify $\Phi\approx   f_\pi^\ast/f_\pi$, which is a consequence of the same $N_c$ dependence of $f_\pi^2$ and $f_\chi^2$ in the CT theory~\cite{CT}. Note that while $g_A$ has no intrinsic density dependence, $f_\pi^\ast$ is directly affected by the IDD because of the locking to $f_\chi^\ast$. This was already noticed in the Adler-Weisberger sum rule.

 At the Fermi-liquid fixed point, the relevant quantities involved are the Landau mass $m_L$,  the Landau interaction parameters  $\tilde{F}_1$ and $\tilde{F}_1^\prime$ and $\Phi=f_\pi^\ast/f_\pi$. Thus the Landau $g^L_A$ -- hence $q^L$ -- must involve only these quantities. The calculation for $g_A^L$ was first done a long time ago~\cite{Friman-Rho}, which to our surprise is exactly reproduced by the considerably more
 improved  argument.
 It is given by
\be
g_A^{\rm L}/ g_A \approx (1-\frac 13 \Phi\tilde{F}_1^\pi)^{-2}\label{gAL}
\ee
where $\tilde{F}_1^\pi$ is the pion Fock term contribution to the Landau parameter $\tilde{F}_1$ that enters in $\delta g_l$. The Fock term is a loop contribution, so naively $O(1/\bar{N})$. But the pion being ``soft," it plays an indispensable role as it does for the anomalous orbital gyromagnetic ratio $\delta g_l^p$.  Since pionic properties are given by chiral dynamics, the pion contribution  $\tilde{F}_1$ can be calculated almost exactly. Thus  once $\Phi$ -- the only parameter in the theory  -- is given, then $g_A^L$ is accurately calculable.   How the pion decay constant behaves in nuclear medium is experimentally measured~\cite{yamazaki}, so $\Phi$ is known in the vicinity of nuclear matter density. Quite surprisingly while $\Phi$ decreases as density increases, the pionic term $\tilde{F}_1^\pi$ increases  with  the product $\Phi\tilde{F}_1^\pi$ staying nearly constant as density changes, say,   between $\sim \frac 12 n_0$ and $\gsim n_0$. Therefore $g_A^L$ is nearly density-independent, which predicts that the quenching factor $q^L$ must be more or less the same from light nuclei to heavy nuclei (and dense matter $n\gsim n_0$). The result is (evaluated at $n_0$)
\be
q^L\simeq 0.79\label{qL}
\ee
that agrees with what's given in $^{100}$Sn~\cite{Sn100},
\be
q_{exp}=0.75(2).
\ee
This implies that the quenching factor $\sim 0.75$ is given  mostly -- if not entirely -- by strong particle-hole correlations affected by neither  ``intrinsic" renormalization via the dilaton condensate nor many-body currents nor higher baryon resonances. We should remark that a similar result was obtained in a $V_{lowk}$-based shell-model approach from mass number $A=48$ to $A=136$~\cite{coraggio}.

\sect{\bf Concluding  Remarks}
The result obtained here exposes the ``old" $g_A$ problem with a totally ``new" face. In nuclei,  $g_A^\ast\approx 1$ is captured entirely by nuclear correlations in the theory with $bs$HLS. In highly dense matter as in compact stars, on the other hand, with the same Lagrangian treated in the limit that the re-parametrized field $\la\chi e^{\frac{i \pi}{f_\pi}}\ra$ goes to zero, called ``dilaton-limit fixed point," which corresponds to the dilaton going massless, $g_A^\ast$ is also found to approach 1~\cite{MR-review}. This can be taken as signaling  the approach to scale invariance. Since the scale symmetry and chiral symmetry are locked, this limit is equivalent to chiral symmetry restoration manifesting in the Adler-Weisberger sum rule. An interesting scenario is that in approaching this limit in conjunction with the vector manifestation, the pion, the dilaton, the $\rho$ (and  $a_1$)  all go massless, giving rise to what we might identify with Weinberg's ``mended symmetry" multiplets~\cite{mended}. At the high density corresponding to this limit, perturbative QCD predicts that the sound velocity in neutron stars should become ``conformal," $(v_s/c)^2=1/3$. Now what's  predicted in  $bs$HLS is surprising: The conformal sound velocity precociously sets in at a density $n\gsim 3n_0$~\cite{MR-review}, far below  what is usually expected, $\gsim 50n_0$. This phenomenon was referred to as ``pseudo-conformal (PC)." This suggests that $g_A^\ast\to 1$ at $n\gsim 3n_0$. We interpret this as signaling the emergence of  the hidden scale symmetry  in baryonic medium. This observation is consistent with  that in pionless EFT,  with all mesons -- including the pion -- {\it integrated out},  the {\it unitarity limit} is applicable in nuclei and compact stars at low energy~\cite{vankolck} and in $bs$HLS with all mesons {\it integrated in}, the pseudo-conformality is applicable in compact stars at high density. 

The author is grateful for discussions with and comments from Yong-Liang Ma.




%

\end{document}